\documentclass[10pt,conference]{IEEEtran}

\usepackage{amsmath,amsfonts}
\usepackage{algorithm}
\usepackage{array}
\usepackage[caption=false,font=normalsize,labelfont=sf,textfont=sf]{subfig}
\usepackage{textcomp}
\usepackage{stfloats}
\usepackage{url}
\usepackage{verbatim}
\usepackage{graphicx}
\usepackage{cite}
\usepackage{graphicx} 
\usepackage{amsmath}
\usepackage{amsmath}
\usepackage{amsmath, amssymb}
\usepackage{xcolor}	
\usepackage{color}
\usepackage{amsmath} 

\usepackage{algpseudocode}
\usepackage{graphics}
\usepackage{amsmath, amsthm}

\allowdisplaybreaks
\usepackage[caption=false,font=normalsize,labelfont=sf,textfont=sf]{subfig}

\usepackage{multicol,lipsum}

\begin{document}

 \title{Robust Resource Allocation for Over-the-Air Computation Networks with Fluid Antenna Array}








\author{\IEEEauthorblockN{Saeid Pakravan\IEEEauthorrefmark{1}, Mohsen Ahmadzadeh\IEEEauthorrefmark{2}, Ming Zeng\IEEEauthorrefmark{1}, Ghosheh Abed Hodtani\IEEEauthorrefmark{2}, Jean-Yves Chouinard\IEEEauthorrefmark{1}, \\and Leslie A Rusch\IEEEauthorrefmark{1}}

\IEEEauthorblockA{\IEEEauthorrefmark{1}Department of Electric and Computer Engineering, Laval University, Quebec, Canada}

\IEEEauthorblockA{\IEEEauthorrefmark{2}Department of Electric and Computer Engineering, Ferdowsi University of Mashhad, Mashhad, Iran}

\IEEEauthorblockA {Email: saeid.pakravan.1@ulaval.ca; m.ahmadzadehbolghan@mail.um.ac.ir; ming.zeng@gel.ulaval.ca; hodtani@um.ac.ir; \\Jean-Yves.Chouinard@gel.ulaval.ca; leslie.rusch@gel.ulaval.ca}}

\maketitle

\begin{abstract}
Fluid antenna (FA) array is envisioned as a promising technology for next-generation communication systems, owing to its ability to dynamically control the antenna locations.   
In this paper, we apply FA array to boost the performance of over-the-air computation networks. Given that channel uncertainty will impact negatively not only the beamforming design but also the antenna location optimization, robust resource allocation is performed to minimize the mean squared error of transmitted messages. 
Block coordinate descent is adopted to decompose the formulated non-convex problem into three subproblems, which are iteratively solved until convergence. Numerical results show the benefits of FA array and the necessity of robust resource allocation under channel uncertainty.
\end{abstract}
	

\section{Introduction}\thispagestyle{empty}

Artificial intelligence advancements have ushered in a new era of intelligent services, creating unprecedented demands for extensive connectivity and vast data processing capabilities. However, meeting these demands is challenging due to limited radio resources and stringent latency requirements. To address these challenges, over-the-air computation (AirComp) has emerged as a transformative technology [1], [2]. AirComp represents a fundamental shift by seamlessly integrating communication and computation processes through the superposition property of multiple-access channels. Unlike conventional methods that sequence computation after communication, AirComp enables simultaneous computation during communication. This innovation leads to more efficient spectrum utilization and facilitates low-latency wireless data aggregation, allowing multiple devices to concurrently use the same channel for transmission. Consequently, AirComp not only enhances spectral efficiency but also addresses the pressing need for real-time data processing in modern intelligent systems.

The integration of wireless communication with computational tasks in AirComp has been extensively investigated in prior research to enable edge processing. Initial studies focused on theoretical frameworks and algorithmic approaches for edge computation over wireless channels, emphasizing the benefits of offloading computational tasks from resource-constrained devices to more powerful remote servers [3]. Subsequent research explored practical implementations and performance evaluations of AirComp systems across various application scenarios, including wireless sensor networks and mobile edge computing platforms [4]. These studies examined factors such as communication latency, energy efficiency, and computational complexity to optimize overall system performance. Furthermore, recent investigations have explored the integration of machine learning and artificial intelligence techniques to enhance intelligent decision making and data analytics within AirComp systems [5], [6]. Furthermore, there is growing interest in combining AirComp with reconfigurable intelligent surfaces (RIS) to further enhance distributed computation capabilities and enable advanced applications in dynamic environments [7-9].

Beyond RISs, the emergence of fluid antenna (FA) systems has introduced novel avenues for manipulating wireless channel conditions, offering new degrees of freedom (DoFs) compared to fixed-position antennas (FPAs) [10-13]. FAs, inspired by the fluidic behavior of electromagnetic fields, provide unprecedented flexibility and adaptability in shaping the radiation patterns and beamforming capabilities of wireless communication systems. By dynamically adjusting the properties of the electromagnetic field, FAs enable efficient spatial multiplexing and beamforming, thereby enhancing the capacity and reliability of wireless communication links. Recently, some paper have highlighted the potential applications of FAs in AirComp networks [14], [15]. However, despite the promising potential of FA systems, their integration into AirComp architectures and optimization under channel uncertainty remain underexplored. Since the optimization of beamforming and FA array location directly depends on the knowledge of channel state information (CSI), addressing this issues is paramount for realizing the full potential of FA systems within AirComp frameworks. 

We address this gap by investigating the optimization of mean squared error (MSE) for an AirComp system that utilizes FA under channel uncertainty. Our research aims to develop novel transceiver designs that mitigate the impact of channel estimation errors, thereby enhancing the robust performance of AirComp systems. Through analytical insights and numerical simulations, we demonstrate the efficacy of our proposed design and their superiority over existing benchmark, ultimately advancing the understanding and practical implementation of AirComp in modern wireless communication scenarios.

{\bf{Notation:}} We employ the following notation system. Scalars are denoted by standard lowercase letters, while vectors are represented using bold lowercase letters. The sets of real numbers and complex numbers are represented by $\mathcal{R}$ and $\mathcal{C}$, respectively. The transposition operation is indicated by $(\cdot)^T$, while the conjugate transpose is represented by $(\cdot)^H$. The $i$-th element in vector $\boldsymbol{x}$ is denoted as $x_i$, while $\left\| \boldsymbol{x} \right\|$ signifies its $\ell_2$-norm. Furthermore, $\odot$ represents the Hadamard product, $\nabla$ symbolizes the gradient operator, $\boldsymbol{I}$ denotes the identity matrix, and $\mathbb{E}$ signifies the expectation operator.


\section{System Model and Problem Formulation}

 As depicted in Fig. 1, we consider a system with $K$ single-antenna users and an access point (AP) with $N$ antennas. The data generated by user $k \in \{1, ..., K\}$, are denoted as $s_k \in \mathcal{C}$, satisfying $\mathbb{E}[s_k] = 0$ and $\mathbb{E}[|s_k|^2] = 1$. Additionally, we assume that the data stream from different users are uncorrelated, meaning $\mathbb{E}[s_k s_j^*] = 0$ for $k \neq j$. The objective of AirComp is to aggregate the user data at the AP, denoted as $s = \frac{1}{K}\sum_{k=1}^{K} s_k$, by exploiting the superposition property of the wireless multiple-access channel. The AP utilizes an adaptable array of FAs, where $N$ FAs can be positioned along a one-dimensional line segment of length $L$. Each FA's position is denoted by $x_n \in [0, L]$, and collectively, the vector $\boldsymbol{x} = [x_1, \ldots, x_N]^T$ represents the antenna position vector (APV), subject to the constraint $0 \leq x_1 < \ldots < x_N \leq L$.
 
 Under line-of-sight (LoS) propagation conditions, the channel coefficient from user $k$ to the AP, denoted as $\boldsymbol{h}_k \in \mathcal{C}^{N \times 1}$, can be expressed as [16]:
\begin{equation}
\boldsymbol{h}_k =\sqrt{{l_k^{-\alpha}}} \begin{bmatrix}
    e^{j \frac{2\pi}{\lambda} x_1 \cos(\theta_k)}, \ldots, e^{j \frac{2\pi}{\lambda} x_N \cos(\theta_k)}
\end{bmatrix}^T,
\end{equation}
where $l_k$ denotes the distance between the user $k$ and AP, and $\alpha$ is the corresponding path loss exponent; both $l_k$ and $\alpha$ are associated with propagation gain. The reception vector in (1) is determined by the APV $\boldsymbol{x}$ and the steering angle $\theta_k$. In this context, $\theta_k$ represents the angle of arrival (AoA) of the LoS path and $\lambda$ represents the wavelength. This equation highlights the dynamic nature of the channel coefficient in FA systems, where it adapts in real-time to optimize signal reception amidst changing environmental conditions and user locations.

Given the set of channel coefficients $\boldsymbol{{h}}_k$, the received signal at the AP can be expressed as
\begin{equation}
\boldsymbol{y} = \sum_{k=1}^{K} \boldsymbol{h}_k b_k s_k + \boldsymbol{z},
\end{equation}
where $b_k$ denotes the transmit coefficient of user $k$, satisfying $\mathbb{E}[|b_k s_k|^2] = |b_k|^2 \leq P_k$, $\forall k \in \mathcal{K}$, where $P_k$ is the maximum power constraint. The additive white Gaussian noise (AWGN) at the AP is represented by $\boldsymbol{z} \in \mathcal{C}^{N \times 1}$. It is characterized by a circularly symmetric complex Gaussian distribution with zero mean and covariance $\sigma_z^2 \boldsymbol{I}$.

For the considered system, it is important to have an accurate estimate of $\boldsymbol h_k$. However, in practical wireless communication systems, the accuracy of channel estimation is affected by various conditions. These conditions, which include multipath propagation, atmospheric variations, physical obstructions and random noise, introduce inevitable uncertainties in the CSI estimation [17-20]. Accurate characterization and knowledge of these uncertainties should be incorporated into system design for several reasons. 

Firstly, the performance of communication systems can be severely degraded if channel uncertainties are ignored. Algorithms and protocols designed under ideal assumptions may be ineffective in real-world scenarios, leading to reduced reliability and efficiency. This becomes even more problematic in the particular case of system based on FA arrays, since both the design of FA array locations and beamforming strategies are affected adversely by channel uncertainty. By accounting for these uncertainties, we can ensure that the system remains robust and adaptive, even under adverse environmental conditions. 

The impact of this uncertainty can be described as an error in phase $\theta_k$ for user $k$ as follows
\begin{equation}
\theta_k=\bar{\theta}_k+\Delta\theta_k,
\end{equation}
where $\bar{\theta}_k$ represents the estimated arrival angle, and $\Delta\theta_k$ represents the uncertainty associated with this estimation. We assume $\Delta\theta_k$ are independent random variables following the uniform distribution over $\left[ -\theta_{k,0}, \theta_{k,0} \right]$, where $\theta_{k,0}$ are constants which quantify the degree of uncertainty. 

Following the methodology outlined in [21], we apply a Taylor expansion to the expression $\exp\left( j \frac{2\pi}{\lambda} x_n \cos(\bar{\theta}_k + \Delta\theta_k) \right)$, and obtain
\begin{multline}
\exp\left( {j \frac{2\pi}{\lambda} x_n\cos(\bar{\theta}_k+\Delta\theta_k)} \right)\approx 
\exp\left( {j \frac{2\pi}{\lambda} x_n \cos(\bar{\theta}_k)} \right)\\+\exp\left( {j \frac{2\pi}{\lambda} x_n \cos(\bar{\theta}_k)} \right)\left( \Delta\theta_k\left( j\frac{2\pi}{\lambda} x_n \sin(\bar{\theta}_k) \right) \right).
\end{multline}
On this basis and after some manipulation, we can re-express the channel modeling in (1) under imperfection as
\begin{equation}
 \boldsymbol{h}_k =\boldsymbol{\bar{h}}_k+\underbrace{\boldsymbol{\bar{h}}_k\odot \boldsymbol{q}(\bar{\theta}_k){\Delta \theta_k}}_{{{\Delta \boldsymbol{h}_k}}}, \ \    \forall k\in \mathcal{K},
\end{equation}
where
\begin{equation}
    \boldsymbol{\bar{h}}_k = {\sqrt{{l_k^{-\alpha}}} \left[ e^{j \frac{2\pi}{\lambda} x_1 \cos(\bar{\theta}_k)}, \ldots, e^{j \frac{2\pi}{\lambda} x_N \cos(\bar{\theta}_k)} \right]^T},
\end{equation}
and
\begin{equation}
\boldsymbol{q}(\bar{\theta}_k)=\sin(\bar{\theta}_k)\left[ j \frac{2\pi}{\lambda} x_1,...,j \frac{2\pi}{\lambda} x_N \right]^T.
\end{equation}
In this context, $\boldsymbol{\bar{h}}_k$ represents the estimated channel vector for user $k$, while $\Delta \boldsymbol{h}_k$ signifies the channel estimation error, modeled as a random vector, with zero mean and diagonal covariance matrix, whose $k$-th diagonal term is given by $\psi_k x_n^2 \theta_{k,0}^2$, where, $\psi_k=\frac{1}{3}(\frac{2\pi}{\lambda} \sqrt{{l_k^{-\alpha}}}{\text{sin}(\bar{\theta}_k)})^2$.



The AP employs a receive beamforming vector $\boldsymbol{m} \in \mathcal{C}^{N \times 1}$ for aggregating data. Consequently, the received signal is formulated as
\begin{equation}
\boldsymbol{m}^H \boldsymbol{y} = \sum_{k=1}^{K} \boldsymbol{m}^H \left( \boldsymbol{\bar{h}}_k + \Delta \boldsymbol{h}_k \right) b_k s_k + \boldsymbol{m}^H \boldsymbol{z},
\end{equation}
from which the average signal estimate $\hat{s}=\frac{\boldsymbol{m}^H \boldsymbol{y}}{K}$ is derived.

To assess the AirComp performance, we use the MSE between the estimated signal $\hat{s}$ and the true average signal $s$, i.e.,
\begin{multline}
\text {MSE}=\mathbb{E}\left[ \left| \hat{s}-s  \right|^2 \right]=\frac{1}{K^2}
\sum_{k=1}^{K}\bigg(\left| {\boldsymbol{m}^H \boldsymbol{ \bar{h}}_k b_k}-1 \right|^2+
\\ \sum_{n=1}^{N}\left| b_k \right|^2\psi_k \theta_{k,0}^2\left| m_n x_n \right|^2\bigg)+\frac{\left\| \boldsymbol{m} \right\|^2\sigma_z^2}{K^2},    
\end{multline}
where the expectation is taken over the stochastic variables $\left\{ s_k \right\}$, $\left\{ \Delta h_k \right\}$, and $\boldsymbol{z}$. The \text {MSE} is composed of three components: the signal misalignment error, the error caused by noise, and the error resulting from inaccuracies in CSI.

The objective of our study is to minimize the MSE described in equation (9) by optimizing the $\boldsymbol{b} = [b_1, \ldots, b_k]^T$, the receive beamforming vector $\boldsymbol{m}$ at the AP, and the FA array location vector $\boldsymbol{x}$, while adhering to the power constraint of each user. The minimization problem for computing MSE in the presence of channel estimation errors is formulated as follows, with the omission of the constant coefficient $\frac{1}{K^2}$ in (9) to simplify the notation.
\begin{equation}
		\begin{aligned}
		&\min_{\boldsymbol{b}, \boldsymbol{m}, \boldsymbol{x}} \hspace{0.2cm} 
 \text{MSE} && \\	&\text{s.t. } C_1:   \left| b_k \right|^2\leq P_k  , \ \ \forall k \in \mathcal{K}, \\
         & \ \ \ \ C_2:x_1\ge 0 ,\ x_N\le L,\\
         & \ \ \ \ C_3:x_n-x_{n-1}\ge L_0, \ \ \forall n=2,...,N.\\		     
	\end{aligned}
\end{equation}
 Here, constraint $C_1$ ensures that each user's transmission power does not exceed its maximum allowable power. Constraint $C_2$ restricts the positioning of FAs within the feasible interval $[0, L]$. Finally, constraint $C_3$ specifies that the distance between any two consecutive FAs is at least $L_0$, thereby avoiding antenna coupling issues. Achieving this globally optimal solution is highly challenging, particularly due to the nonconvex nature of the objective function and the coupling of the variables.

\begin{figure}[ht]
    \centering
    \includegraphics [width=7.5cm, height=3.9cm]{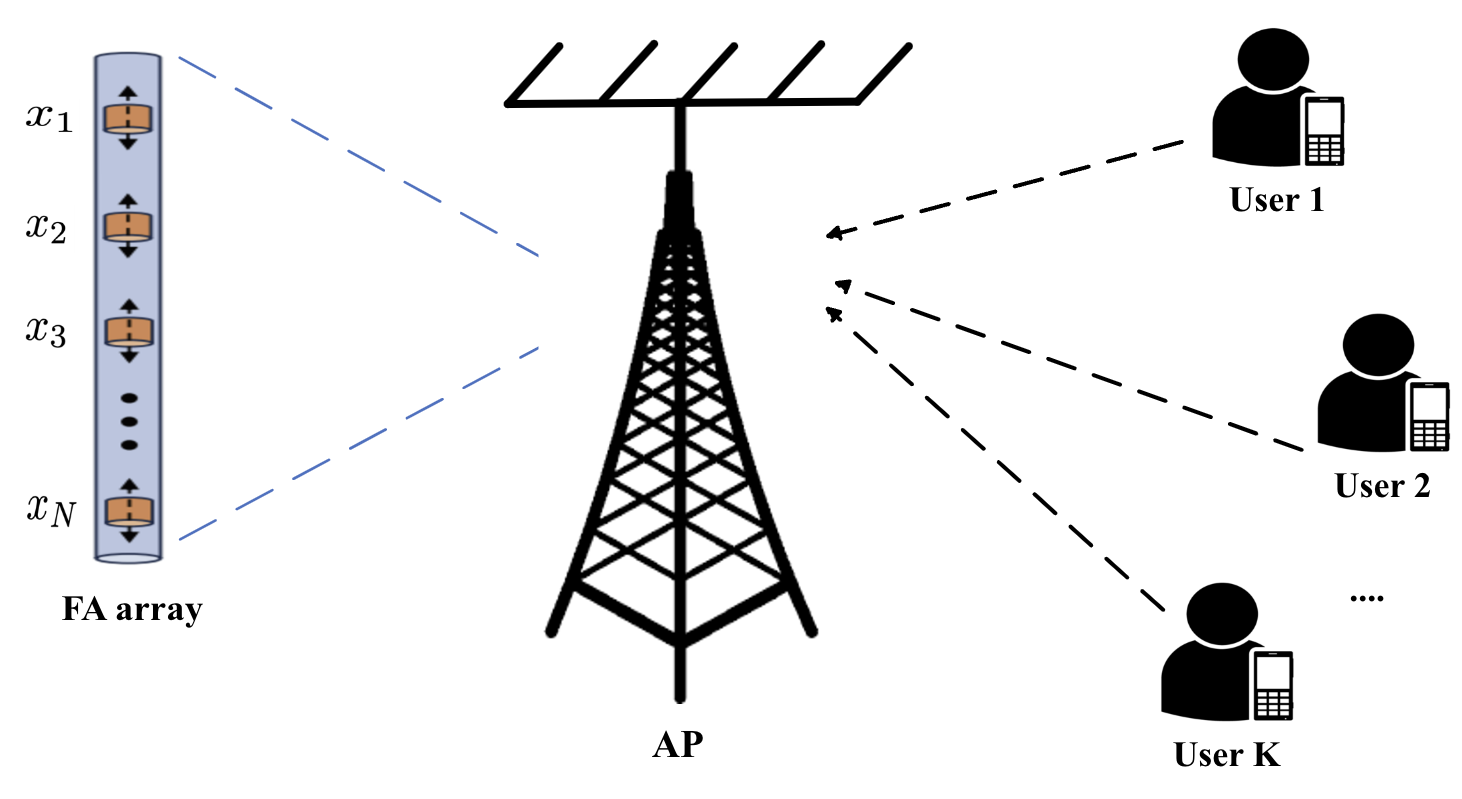}%
    \qquad
    \caption{An illustration of the system model.}%
    \label{fig:example}%
\end{figure}

\section{Proposed Solution}

In this section, we utilize the concept of block coordinate descent, which involves sequentially optimizing certain variables while holding the remaining variables constant.

\subsection{Subproblem 1: Transmit power allocation}

 First, we optimize $\boldsymbol{b}$ under given other parameters. The associated optimization problem with respect to $\boldsymbol{b}$ can be formulated as follows:
\begin{equation}
		\begin{aligned}
		&\min_{\boldsymbol{b}} \hspace{0.2cm} 
\sum_{k=1}^{K}\left( \left| {\boldsymbol{m}^H \boldsymbol{ \bar{h}}_k b_k}-1 \right|^2+
 \left| b_k \right|^2\psi_k \theta_{k,0}^2 \sum_{n=1}^{N}\left| m_n x_n \right|^2 \right)&& 
  \\
		&\text{s.t. } \ \   \left| b_k \right|^2\leq P_k  , \ \ \forall k \in \mathcal{K}. \\
	\end{aligned}
\end{equation}
We note that \( b_1, \ldots, b_K \) in (11) are independent and can be separated into \( K \) individual subproblems. The subproblem corresponding to each \( b_k \), for each \( k \in K \), is expressed as
\begin{equation}
		\begin{aligned}
		&\min_{b_k} \hspace{0.2cm} 
\left| {\boldsymbol{m}^H \boldsymbol{ \bar{h}}_k b_k}-1 \right|^2+
 \left| b_k \right|^2\psi_k \theta_{k,0}^2 \sum_{n=1}^{N}\left| m_n x_n \right|^2 && 
  \\
		&\text{s.t. } \  \ \left| b_k \right|^2\leq P_k. \\
	\end{aligned}
\end{equation}
Upon examining the derivative of the objective function within the subproblem, we derive the optimal solution for $b_k$:
\begin{equation}
  b_k^{\star}=\text {min}\left( \sqrt{P_k},\frac{\left| \boldsymbol{m}^H \bar{\boldsymbol{h}}_k\right|}{\left| \boldsymbol{m}^H \bar{\boldsymbol{h}}_k\right|^2+
\psi_k \theta_{k,0}^2 \sum_{n=1}^{N}\left| m_n x_n \right|^2
} \right).
\end{equation}

\subsection{Subproblem 2: Receiver beamforming design}

Following this, we proceed to optimize \( \boldsymbol{m} \) in (10) with the $\boldsymbol{b}$ and $\boldsymbol{x}$ fixed. This involves solving the following unconstrained convex optimization problem:
\begin{equation}
		\begin{aligned}
		&\min_{\boldsymbol{m}} \hspace{0.2cm} 
\text{MSE}. 
  \\	\end{aligned}
\end{equation}
To solve this optimization problem, we take the gradient of the objective function and set it equal to zero. This allows us to derive the optimal solution for the $\boldsymbol{m}$, which is given by:
\begin{equation}
{\boldsymbol m}^{\star}=R^{-1}\sum_{k=1}^{K}\bar{\boldsymbol{h}}_k b_k,
\end{equation}
where $R=\sigma_z^2 \boldsymbol{I}+\sum_{k=1}^{K}\left| b_k \right|^2( \bar{\boldsymbol{h}}_k \bar{\boldsymbol{h}}_k^H+ \psi_k xx^H \theta_{k,0}^2\boldsymbol{I})$. The optimized receive beamforming solution outlined in (15) exhibits a sum-minimum MSE structure. This summation approach is intuitive as it aggregates signals more effectively from all users, thereby facilitating functional computation.

\subsection{Subproblem 3: FA array design}

Finally, the optimization problem associated with $\boldsymbol{x}$ is expressed as follows:
\begin{equation}
		\begin{aligned}
		&\min_{\boldsymbol{x}} \hspace{0.2cm} 
\sum_{k=1}^{K}\left( \left| {\boldsymbol{m}^H \boldsymbol{ \bar{h}}_k b_k}-1 \right|^2+
 \left| b_k \right|^2\psi_k \theta_{k,0}^2 \sum_{n=1}^{N}\left| m_n x_n \right|^2 \right) && 
  \\&\text{s.t. } C_2 \ \text {and}\ C_3. \end{aligned}
\end{equation}
This objective function is highly non-convex. To address it, we employ the Broyden-Fletcher-Goldfarb-Shanno (BFGS) method to obtain a locally optimal $\boldsymbol{x}$. The BFGS algorithm belongs to the class of quasi-Newton methods and is particularly popular due to its robustness and efficiency in practice [22], [23]. The Appendix provides a detailed explanation of this algorithm for solving the mentioned problem.

Algorithm 1 outlines the approach for designing the proposed transceiver and APV. Its convergence is guaranteed since in each iteration, the objective function declines or remains unchanged.

\begin{algorithm}
\caption{Pseudo-code for the proposed design of transceivers and APV}
\begin{algorithmic}[1]
    \State \textbf{Initialization:}
    \State \quad Set initial power allocation coefficients: \( b_k = \sqrt{P_k} \), \( \forall k \in \mathcal{K} \)
    \State \quad Set initial APV positions: \( x_n = \frac{L n}{N+1} \), \( \forall n \in \{1, \ldots, N\} \)
    \State \quad Set convergence tolerance: \( \epsilon \)
    \State \quad Initialize iteration counter: \( i = 0 \)
    \While{not converged}
        \State \quad \( i \gets i + 1 \)
        \State \quad \textbf{Step 1: Update Receive Beamforming Vectors}
        \State \quad \quad Update \( \boldsymbol{m}_{i} \) using equation (15)
        \State \quad \textbf{Step 2: Update Power Allocation Coefficients}
        \State \quad \quad Update \( \boldsymbol{b}_{i} \) by solving equation (13)
        \State \quad \textbf{Step 3: Update APV Positions}
        \State \quad \quad Update \( \boldsymbol{x}_{i} \) using the BFGS method
        \State \quad \textbf{Step 4: Check Convergence}
        \State \quad \quad If \( \| \boldsymbol{m}_{i} - \boldsymbol{m}_{i-1} \| < \epsilon \) and \( \| \boldsymbol{b}_{i} - \boldsymbol{b}_{i-1} \| < \epsilon \) and \( \| \boldsymbol{x}_{i} - \boldsymbol{x}_{i-1} \| < \epsilon \), then converge
    \EndWhile
\end{algorithmic}
\end{algorithm}

\section{Simulation Results}

This section assesses the effectiveness of our proposed AirComp design based on computation the MSE performance. The simulation setup includes the following parameters: the initial feasible interval \(L_0 = 0.5\lambda\), whereas the maximum feasible interval \(L = 8\lambda\); Without prejudice to the conclusion, $\lambda$ is set to 1 for simplification, while equal maximum power constraint for all users \(P_k = P_0\), \(\forall k \in \mathcal{K}\). Additionally, for simplicity, we assume the same variance for CSI errors among all users, i.e., $\theta_{k,0}=\theta_0$, for all $k$. We compare the performance of our proposed designs with two benchmark schemes: one that ignores CSI errors and another that employs FPA. For the benchmark that ignores CSI errors, the AP and users optimize the transceiver design by solving the optimization problem without accounting for channel estimation errors. For the FPA benchmark, the positions of \( N \) FAs are uniformly distributed across the feasible region \([0, L]\). This setup fixes the APV as 
\( \boldsymbol{x} = \left[\frac{L}{N+1}, \cdots, \frac{NL}{N+1}\right]^{T} \). The optimal configurations for \( \boldsymbol{b} \) and \( \boldsymbol{m} \) are determined iteratively using (13) and (15) until convergence is achieved.

\begin{figure}[ht]
    \centering
    \includegraphics [width=7.5cm, height=5cm]{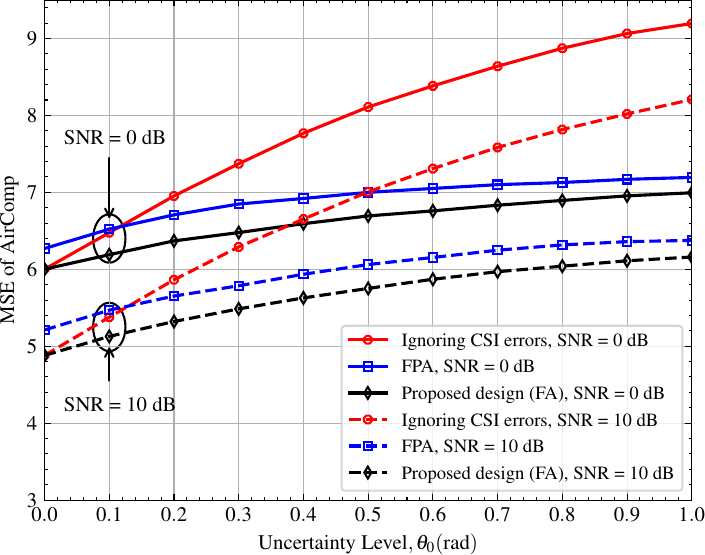}%
    \qquad
    \caption{Computation of MSE versus the uncertainty level under different SNR levels, where $N=8$ and $K=10$.}%
\end{figure}

Fig. 2 illustrates the computation of MSE relative to the uncertainty level, for different signal-to-noise ratio (SNR) levels, with \( K = 10 \) and \( N = 8 \). It is noteworthy that throughout the range of $\theta_{0}$, the proposed design consistently outperforms alternative benchmarks for both SNR settings. Moreover, as $\theta_{0}$ increases, the performance gap between the proposed design and the benchmark that ignores CSI widens significantly. This superiority highlights the robustness of our approach, which maintains better performance even in the presence of significant uncertainties in CSI.

\begin{figure}[ht]
    \centering
    \includegraphics [width=7.5cm, height=5cm]{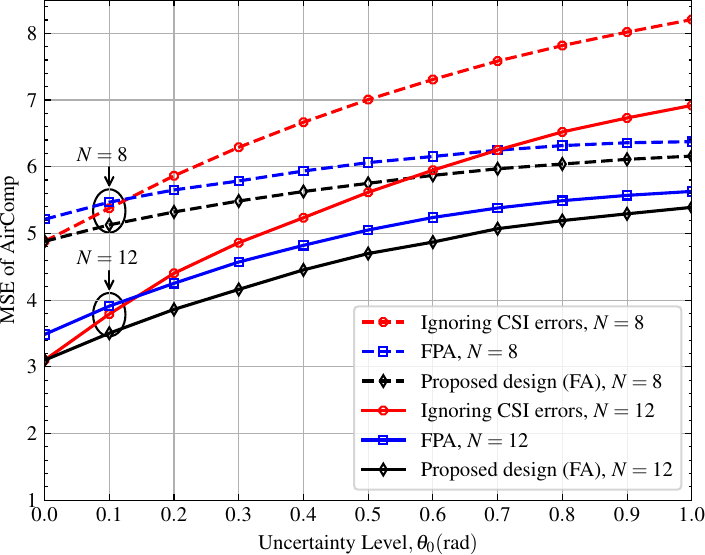}%
    \qquad
    \caption{Computation of MSE versus the uncertainty level under different numbers of FAs, where $\text{SNR}=10 \text{dB}$ and $K=10$.}%
\end{figure}

Fig. 3 shows the performance of the MSE computation for the aforementioned scheme versus the uncertainty level, under different numbers of FAs, i.e., \(N\). As expected, MSE performance improves (i.e., decreases) consistently for all schemes as the number of FAs increases. This behavior is expected because having more FAs allows for better spatial diversity and improved signal aggregation. Notably, our proposed scheme exhibits a significant performance advantage over the FPA benchmark at both \(N = 8\) and \(N = 12\), underscoring the benefits of optimizing the positions of FAs.

\begin{figure}[ht]
    \centering
    \includegraphics [width=7.5cm, height=5cm]{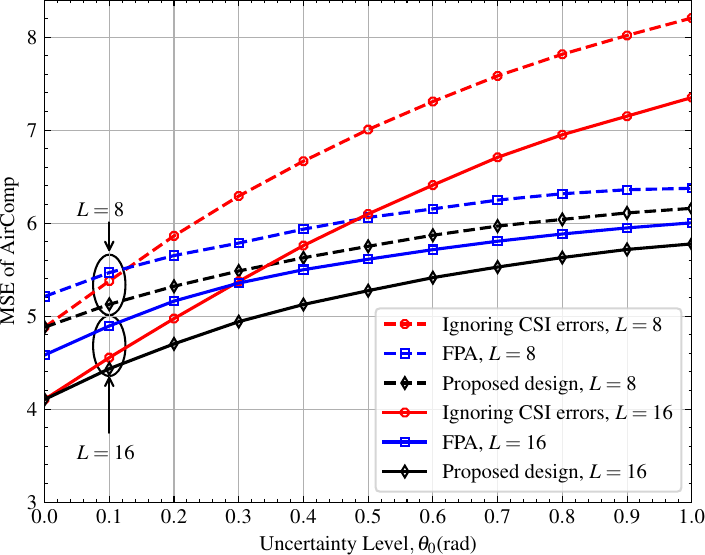}%
    \qquad
    \caption{Computation of MSE versus the uncertainty level under different values of the feasible interval $L$, where $\text{SNR}=10 \text{dB}$ and $N=8$.}%
\end{figure}

Fig. 4 illustrates the MSE computation versus the uncertainty level for different values of the feasible interval $L$, with $\text{SNR} = 10\,\text{dB}$ and $N = 8$. The results indicate that as $L$ increases, the MSE performance improves for all the considered schemes. Note that the performance gap between the proposed scheme and FPA benchmark enlarges with $L$, especially when channel uncertainty is minor. This improvement can be attributed to the improved flexibility in FA positioning.

\section{Conclusion}

In this paper, we conducted a comprehensive performance analysis of AirComp networks integrated with FA arrays in the presence of channel uncertainty. The focus of our study was to jointly optimize the transceiver design and APV to minimize MSE. We addressed this nonconvex optimization problem by using the block coordinate descent technique, which decomposed the problem into three manageable subproblems that were iteratively solved to achieve convergence. Numerical results showed that our proposed transceiver and APV design obtains notable improvements in MSE performance compared to traditional FPA arrays, and its better resilience against channel uncertainty.







\appendix

We employ the BFGS method to address our optimization problem, effectively incorporating constraints into the objective function through a barrier term. This quasi-Newton method is particularly efficient for managing constraints. The barrier function is defined to accommodate the constraints of our problem. Specifically, we have constraints on the range of values for \(x_1\) and \(x_N\), and constraints that control the distance between each consecutive pair of \(x_n\):
\begin{equation}
    \Phi(\boldsymbol{x}) = \log(x_1) + \log(L - x_N) + \sum_{n=2}^N \log(x_n - x_{n-1} - L_0).
\end{equation}
The overall objective function, which includes the barrier function, is defined as follows:
\begin{multline}
     L(\boldsymbol{x}, \mu) = \sum_{k=1}^{K}(\left| {\boldsymbol{m}^H \bar{\boldsymbol{h}}_k b_k}-1 \right|^2+\\\left| b_k \right|^2\psi_k \theta_{k,0}^2 \sum_{n=1}^{N}\left| m_n x_n \right|^2) + \mu \Phi(\boldsymbol{x}).
\end{multline}
To compute the updates required for the BFGS method, we approximate the gradient and the Hessian matrix of the function \(L(\boldsymbol{x}, \mu)\). The BFGS method is used to iteratively update the estimate of the optimal point. The overall procedures of the BFGS method are outlined as follows.

\begin{algorithm}
\label{algorithm2}
\caption{Pseudo-code for solving (16) with the BFGS method}
\begin{algorithmic}[1]
\State Choose initial values \( \boldsymbol{x} \) satisfying \( x_1 > 0 \), \( x_N < L \), and \( x_n - x_{n-1} > L_0 \) for all \( n = 2, \ldots, N \).
\State Initialize parameters: \( \mu > 0 \), tolerance levels \( \epsilon_{\text{grad}} \), \( \epsilon_{\text{step}} \), and initial positive-definite approximation to the Hessian \( H = I \).
\State \( \boldsymbol{x}_{\text{prev}} \leftarrow \boldsymbol{x} \)
\While{ \( \| \nabla L(\boldsymbol{x}, \mu) \| \geq \epsilon_{\text{grad}} \) \textbf{or} \( \| \boldsymbol{x}_{\text{new}} - \boldsymbol{x} \| \geq \epsilon_{\text{step}} \)}
    \State Compute the barrier function \( \Phi(\boldsymbol{x}) \).
    \State Compute the overall objective function \( L(\boldsymbol{x}, \mu) \).
    \State Compute the gradient \( \nabla L(\boldsymbol{x}, \mu) \).
    \State Compute the search direction \( p = -H \nabla L(\boldsymbol{x}, \mu) \).
    \State Conduct a line search to find an appropriate step size \( \alpha \).
    \State Update \( \boldsymbol{x}_{\text{new}} = \boldsymbol{x} + \alpha p \).
    \State Compute the gradient at \( \boldsymbol{x}_{\text{new}} \), \( \nabla L(\boldsymbol{x}_{\text{new}}, \mu) \).
    \State Compute \( s = \boldsymbol{x}_{\text{new}} - \boldsymbol{x} \) and \( y = \nabla L(\boldsymbol{x}_{\text{new}}, \mu) - \nabla L(\boldsymbol{x}, \mu) \).
    \State Update the Hessian approximation \( H \) using the BFGS formula:
    \[
    H \leftarrow H + \frac{yy^T}{y^T s} - \frac{H s s^T H}{s^T H s}
    \]
    \State Evaluate \( L(\boldsymbol{x}_{\text{prev}}, \mu) \) and \( L(\boldsymbol{x}_{\text{new}}, \mu) \).
    \If{ \( L(\boldsymbol{x}_{\text{new}}, \mu) < L(\boldsymbol{x}_{\text{prev}}, \mu) \) }
        \State \( \boldsymbol{x} \leftarrow \boldsymbol{x}_{\text{new}} \)
    \Else
        \State \( \boldsymbol{x} \leftarrow \boldsymbol{x}_{\text{prev}} \)
    \EndIf
    \State \( \boldsymbol{x}_{\text{prev}} \leftarrow \boldsymbol{x} \)
    \State Reduce \( \mu \) (e.g., \( \mu \leftarrow \mu / \tau \) for some \( \tau > 1 \)).
\EndWhile
\end{algorithmic}
\end{algorithm}

This process is iterated by gradually reducing \(\mu\) to approach the boundary of the constraints and converge to the local optimal point.


	\bibliographystyle{IEEEtran}
	\bibliography{maindocument}

	\section{References Section}


	\vfill
	
\end{document}